\documentclass[epj]{webofc}
\usepackage[varg]{txfonts}   
\usepackage{amsmath}
\usepackage{slashed}
\usepackage{amstext}
\usepackage{amssymb}
\usepackage{bm}

\wocname{EPJ Web of Conferences}
\woctitle{CONF12}
\def\bea{\begin{eqnarray}}
\def\eea{\end{eqnarray}}

\newcommand{\sigp}{{\bm \sigma}\cdot{\hat{\bf p}}}

\newcommand{\xip}{{\bm \xi}\cdot{\hat{\bf p}}}

\newcommand{\sigxi}{{\bm \sigma}\cdot{\hat{\bm \xi}}}

\begin{document}
\selectlanguage{english}
\title{Relativistic phenomenology of meson spectra with a covariant quark model in Minkowski space}
%
%

\author{Sofia Leit\~ao\inst{1}\fnsep\thanks{\email{sofia.leitao@tecnico.ulisboa.pt}} \and
        Alfred Stadler\inst{2} \and
        M. T. Pe\~na \inst{3,1}  \and
        Elmar P. Biernat \inst{1}
}

\institute{CFTP, Instituto Superior T\'ecnico, Universidade de Lisboa, Av. Rovisco Pais 1, 1049-001 Lisboa, Portugal 
\and
           Departamento de F\'isica, Universidade de \'Evora, 7000-671 \'Evora, Portugal 
\and
           Departamento de F\'isica, Instituto Superior T\'ecnico, Universidade de Lisboa, Av. Rovisco Pais 1, 1049-001 Lisboa, Portugal
}

\abstract{In this work, we perform a covariant treatment of quark-antiquark systems.\,We calculate the spectra and wave functions using a formalism based on the Covariant Spectator Theory (CST). Our results not only reproduce very well the experimental data with a very small set of global parameters, but they also allow a direct test of the predictive power of covariant kernels.}
\maketitle
\section{Introduction}
\label{intro}  

A thorough description of the properties of all observed hadrons cannot yet be derived directly from QCD.\,Despite this fact, in the past few decades, the problem of strongly-bound systems has been studied successfully in a complementary way by a panoply of different approaches.\,They can be cast roughly into three categories~\cite{Brambilla}: effective field theories (growing out of operator-product expansions and the formalism of phenomenological Lagrangians), lattice gauge theories (the discretized version of QCD), and other nonperturbative approaches.\,In the last category, among the most used approaches are: large $N_c$, generalizations of the Shifman-Vainstein-Zakhrov sum rules, QCD vacuum models and effective string models, the AdS/CFT conjecture, and the Schwinger-Dyson/Bethe-Salpeter equations.

Our approach, the Covariant Spectator Theory (CST) \cite{Gro69}, is a quantum field theoretical formalism similar to the Schwinger-Dyson/Bethe-Salpeter method. The main idea of the CST is to turn the kernel of the four-dimensional Bethe-Salpeter equation into an equivalent form, with a different two-body propagator and an accordingly changed interaction kernel.\,The new propagator is chosen such that the original four-dimensional integration reduces to a three-dimensional integration while the manifest covariance of the equation is maintained.

The CST prescription for the two-body propagator is motivated by partial cancellations that occur between the Bethe-Salpeter ladder and crossed-ladder diagrams, with a net result that is close to the CST ladder diagrams only. This amounts to a very efficient way of summing the Bethe-Salpeter series, which was shown in the application of CST to nucleon-nucleon scattering leading to a high precision NN potential with a reduced number of parameters~\cite{Gro08}.

The details of the CST applied to mesons can be found in Refs.~\cite{Gross:1991te, GMilana:1994,Uzzo,CST:2014}. In the present work we report on the results of heavy quarkonium and heavy-light $q\bar{q}$ states for pseudoscalar, scalar, vector and axial-vector mesons. The CST formalism is particularly suitable for the treatment of heavy-light and heavy systems, since the CST equation reduces in the one-body and nonrelativistic limits to the Dirac and the Schr\"odinger equations, respectively.  

\section{Model and numerical implementation}
\label{model}  
\subsection{Equation for the vertex function}
\begin{figure}[h!]
\centering
\includegraphics[width=8cm,clip]{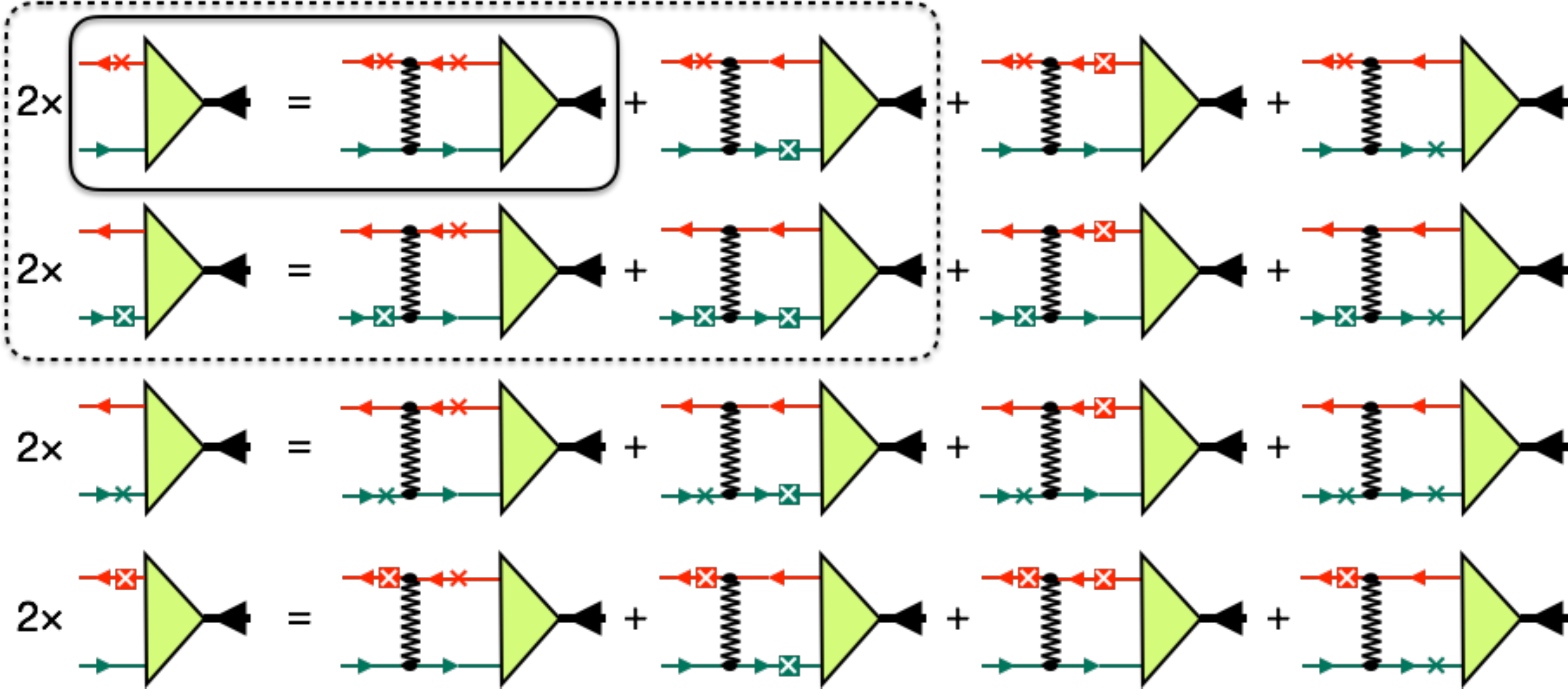}
\caption{The four-channel CST equation (4CSE). The solid rectangle indicates the one-channel CST equation (1CSE) used in this work, the dashed rectangle a two-channel extension with charge-conjugation symmetry. Crosses on quark lines indicate that only the positive-energy pole contribution of the propagator is kept, light crosses in a dark square refer to the negative-energy pole contribution.}
\label{fig}       
\end{figure}

The starting point to derive the CST equation is the Bethe-Salpeter (BS) equation for the quark anti-quark vertex function $\Gamma_{BS}(p_1,p_2)$  with an irreducible interaction kernel ${\cal V}(p,k;P) $ ($P$ is the two-body total 4-momentum, and $p$ and $k$ are the external and internal relative 4-momenta, respectively), given by
\begin{equation}
\Gamma_{BS}(p_1,p_2)= i\int\frac{ {d}^4k}{(2\pi)^4}\,{\cal V}(p,k;P) 
S_1({k}_1)\,\Gamma_{BS}(k_1,k_2)\,S_2(k_2)\,,
\label{eq:BS}
\end{equation}
where $S_i(k_i)$ is the dressed quark propagator depending on the individual 4-momentum $k_i$ of quark $i$. The CST prescription described in the introduction is to assume particle 1, the heaviest particle with mass $m_1$, to be on its mass-shell.\,This yields the CST equation for the vertex function $\Gamma_{1CS}$, where \lq\lq 1CS" or \lq\lq 1CSE" stands for \textit{one-channel} spectator equation \cite{CST:2014}. Specifically, the 1CSE results from the BS by taking into account only the contribution from the residue of the pole that appears when particle $1$ is placed on its positive-energy mass-shell. More contributions could also be included which leads to a coupled set of CST equations depicted diagrammatically in Fig.\,\ref{fig}.\,However, for the heavy and heavy-light systems the 1CSE is a good approximation \cite{Leitao:2016bqq}, as it retains the most important properties of the complete set of CST equations, namely manifest covariance, cluster separability, and the correct one-body limit.\,It is also a good approximation for equal-mass particles, as long as the bound-state mass is not too small. However, a property the 1CSE does not maintain is charge-conjugation symmetry.\,Therefore, states calculated with the 1CSE are not expected to have a definite C-parity. In principle, this problem is easily remedied by using the two-channel extension inside the dashed rectangle of Fig.\,\ref{fig} instead.
The 1CSE is given by 

\vspace{-0.5cm}
\begin{equation}
\Gamma_{1CS} (\hat{p}_1,p_2)= - \int \frac{d^3k}{(2\pi)^3} \frac{m_1}{E_{1k}} \sum_K V_K(\hat{p}_1,\hat{k}_1) \Theta_{1}^{K(\mu)}  
\frac{m_1+\hat{\slashed{k}}_1}{2m_1} \Gamma_{1CS}(\hat{k}_1,k_2)
\frac{m_2+\slashed{k}_2}{m_2^2-k_2^2-i\epsilon}\Theta^K_{2(\mu)} \, ,
\label{eq:1CS}
\end{equation}
where $\Theta_i^{K(\mu)}={\bf 1}_i, \gamma^5_i,$ or $\gamma_i^\mu$,  $V_K(\hat{p}_1,\hat{k}_1)$ describes the momentum dependence of the interaction kernel $K$, $m_i$ is the mass of quark $i$, and $E_{ik}\equiv (m_i^2+{\bf k}^2)^{1/2}$. A ``$\hat{\phantom{p}}$'' over a momentum indicates that it is on its positive-energy mass shell.

\subsection{Kernel for the interaction}
The interaction kernel consists of a covariant generalization of the linear (L) confining potential used in \cite{Leitao:2014}, a color Coulomb (Coul), and a constant (C) interaction,
\begin{equation}
{\cal V}\equiv
\sum_K V_K  \Theta_{1}^{K(\mu)} \otimes \Theta^K_{2(\mu)}\,= \left[ (1-y) \left({\bf 1}_1\otimes {\bf 1}_2 + \gamma^5_1 \otimes \gamma^5_2 \right) - y\, \gamma^\mu_1 \otimes \gamma_{\mu 2} \right]V_L  -\gamma^\mu_1 \otimes \gamma_{\mu 2} \left[ V_{Coul}+V_C \right]\,.
\label{eq:kernel}
\end{equation}
The mixing parameter $y$  allows to dial between an equal-weighted scalar-plus-pseudoscalar structure, which preserves the chiral-symmetry constraints \cite{CSTpi-pi}, and a vector structure, while preserving the same nonrelativistic limit.\,The precise Lorentz structure of the confining interaction is not known, and by fitting the $y$ parameter from the mesonic spectra some further information can be gained. The momentum-dependent parts of the kernel are given by
\begin{equation}
V_\mathrm{L}(p,k)  = -8\sigma \pi\left[\left(\frac{1}{q^4}-\frac{1}{\Lambda^4+q^4}\right)-\frac{E_{p_1}}{m_1}(2\pi)^3 \delta^3 (\mathbf{q})\int \frac{d^3 k'}{(2\pi)^3}\frac{m_1}{E_{k'_1}}\left(\frac{1}{q'^4}-\frac{1}{\Lambda^4+q'^4}\right)\right],
\end{equation}
\begin{equation}
V_\mathrm{Coul.}(p,k)  = -4 \pi \alpha \left(\frac{1}{q^2}-\frac{1}{q^2-\Lambda^2}\right), \quad
V_\mathrm{C}(p,k) = (2\pi)^3\frac{E_{k_1}}{m_1} C \delta^3 (\mathbf{q})
\label{eq:V}
\end{equation}
where $q^{(\prime)}=p-k^{(\prime)}$.
The three coupling strengths $\sigma$, $\alpha$, and $C$ are free parameters of the model.\,Furthermore, an analysis of the asymptotic behavior for large momenta  $k$ shows that we need to regularize the kernel in order to have convergence. We use Pauli-Villars regularization for both linear and the Coulomb parts, which yields one additional parameter, the cut-off $\Lambda$.

\subsection{Numerical implementation}
In order  to find a numerical solution for the bound-state problem we expand both the projector and the propagator of Eq.\,(\ref{eq:1CS}) in terms of $u^\rho$-spinors (with $\rho=\pm$) defined as follows:
\begin{equation}
 u_i^{+}(p,\lambda)=\sqrt{\frac{E_{i\rho}+m_i}{2m_i}} 
 \left(\begin{array}{c}
   1\\
   \frac{\mathbf{\sigma}\cdot \mathbf{p}}{E_{i p}+m_i}
 \end{array}\right)\otimes\chi_\lambda (\hat{p}),
\qquad u_i^{-}(p,\lambda)=\sqrt{\frac{E_{i p}+m_i}{2m_i}} 
 \left(\begin{array}{c}
   -\frac{\mathbf{\sigma}\cdot \mathbf{p}}{E_{i p}+m_i}\\1
 \end{array}\right)\otimes\chi_\lambda (\hat{p})\,,
\label{eq:spinors}
\end{equation}
and where $\chi_\lambda$ are two-component spinors. Introducing the notation
\begin{equation}
\Theta_{i,\lambda\lambda'}^{\rho\rho' K(\mu) }({\bf p},{\bf k}) \equiv
\bar{u}_i^\rho({\bf p},\lambda) \Theta^{K(\mu)} u_i^{\rho'}({\bf k},\lambda'), \qquad \Gamma^{+\rho'}_{\lambda\lambda'} (p) \equiv \bar{u}_1^+({\bf p},\lambda) \Gamma(p) u_2^{\rho'}({\bf p},\lambda'),
\label{eq:Theta,Gamma}
\end{equation}
for the spinor matrix elements of the interaction vertices and the spinor matrix elements of the vertex function, respectively, we obtain
\begin{align}
\Gamma^{+\rho'}_{\lambda\lambda'} (p)
= 
- \int \frac{d^3k}{(2\pi)^3} \frac{m_1}{E_{1k}}  
\frac{ m_2}{E_{2k}}  
 V(p,k) 
\sum_{\rho\lambda_1 \lambda_2}
\Theta_{1,\lambda\lambda_1}^{++K(\mu)}({\bf p},{\bf k})
\Gamma^{+\rho}_{\lambda_1\lambda_2} (k) 
 \frac{\rho}{E_{2k}-\rho k_{20}}  \Theta_{2,\lambda_2\lambda'K(\mu)}^{\rho\rho'}({\bf k},{\bf p}) \, .
\label{eq:1CSE3}
\end{align}
Multiplying  Eq. (\ref{eq:1CSE3}) from the left by  $\bar{u}_1^+({\bf p},\lambda)$ and from the right by $u_2^{\rho'}({\bf p},\lambda')$  yields
\begin{multline}
\rho'(E_{2p}-\rho' p_{20}) \sqrt{\frac{m_1 m_2}{E_{1p}E_{2p}} }
\rho'\frac{\Gamma^{+\rho'}_{\lambda\lambda'} (p)}{E_{2p}-\rho' p_{20} }
= 
- \int \frac{d^3k}{(2\pi)^3} \sqrt{\frac{m_1 m_2}{E_{1k}E_{2k}} } \sqrt{\frac{m_1 m_2}{E_{1p}E_{2p}} } 
 V(p,k) \\
\times \sum_{\rho\lambda_1 \lambda_2}
\Theta_{1,\lambda\lambda_1}^{++K(\mu)}({\bf p},{\bf k})\sqrt{\frac{m_1 m_2}{E_{1k}E_{2k}} }
\Gamma^{+\rho}_{\lambda_1\lambda_2} (k) 
 \frac{\rho}{E_{2k}-\rho k_{20}}  \Theta_{2,\lambda_2\lambda'K(\mu)}^{\rho\rho'}({\bf k},{\bf p}) \, .
\label{eq:1CSE4}
\end{multline}
By introducing the wave functions when quark 1 is on-shell as
\begin{equation}
 \Psi^{+\rho}_{1,\lambda_1\lambda_2}(\mathbf{k}):=\sqrt{\frac{m_1m_2}{E_{1\mathbf{k}}E_{2\mathbf{k}}}}\frac{\rho}{E_{2\mathbf{k}}-\rho (E_{1\mathbf{k}}-\mu)} \Gamma^{+\rho}_{\lambda_1\lambda_2}(k),
\label{eq:wfs1}
\end{equation}
we can finally cast Eq.\,(\ref{eq:1CS}) into the following form:
\begin{equation}
(\rho' E_{2p}-E_{1p}+\mu) \Psi^{+\rho'}_{1,\lambda\lambda'} (p)
= 
- \int \frac{d^3k}{(2\pi)^3}  N(p,k)
 V(p,k) 
\sum_{\rho\lambda_1 \lambda_2}
\Theta_{1,\lambda\lambda_1}^{++K(\mu)}({\bf p},{\bf k})
\Psi^{+\rho}_{1,\lambda_1\lambda_2} (k) 
 \Theta_{2,\lambda_2\lambda'K(\mu)}^{\rho\rho'}({\bf k},{\bf p}).
\label{eq:1CSE6}
\end{equation}

 In order to proceed we have to specify the Lorentz structure of the vertex function for the meson under study, i.e. a scalar, pseudoscalar, vector, or axial-vector meson. In general, we can always write the wave functions in terms of two-component spinors $\chi$ and  $K_j^\rho(\hat{\bf p})$ operators which are $2\times 2$ matrices, as follows:
\begin{equation}
\Psi^{+\rho}_{1,\lambda\lambda'} (p)=\sum_j \psi_j^\rho(p) \chi^\dagger_\lambda(\hat{\bf p})\, K_j^\rho(\hat{\bf p}) \, \chi_{\lambda'}(\hat{\bf p})\,.
\label{eq:PSI}
\end{equation}
In Table \ref{tab:KK} all the $K_j^\rho(\hat{\bf p})$ used in this work are listed for convenience. 

\begin{table}[h!]
\caption{Wave function components for each meson}
\begin{center}
\begin{tabular}{|l|cc|cc|}
\hline
Meson & $K_1^-(\hat{\bf p})$ & $K_2^-(\hat{\bf p})$ & $K_1^+(\hat{\bf p})$ & $K_2^+(\hat{\bf p})$ \\
\hline
Pseudoscalar      & ${\bf 1}$ & - & ${\bm \sigma}\cdot{\bf \hat p}$ & -\\
Scalar      & ${\bm \sigma}\cdot{\bf \hat p}$ & - &  ${\bf 1}$ & -\\
\hline
Vector      & $\sigxi$ & $\frac{1}{\sqrt{2}}\left(3 \xip\, \sigp-\sigxi\right)$ & $\sqrt{3}\xip$ & $\sqrt{\frac32}\left(\sigxi\, \sigp-\xip\right) $\\
Axial-Vector      &  $\sqrt{3}\xip$ & $\sqrt{\frac32}\left(\sigxi\, \sigp-\xip\right) $ & $\sigxi$ & $\frac{1}{\sqrt{2}}\left(3 \xip\, \sigp-\sigxi\right)$ \\
\hline
\end{tabular}
\end{center}
\label{tab:KK}
\end{table}

The main advantage of using this basis for the wave functions is that it has definite orbital angular momentum and thus enables us to determine the spectroscopic identity of our solutions, which is indispensable when comparing to the measured states. In the nonrelativistic limit, they reduce to the familiar Schr\"odinger wave functions.\, However, our relativistic wave functions contain components not present in nonrelativistic solutions.\,For example, the $S$-waves of our pseudoscalar states couple to small $P$-waves (with opposite intrinsic parity) that vanish in the nonrelativistic limit, whereas, for vector mesons, coupled $S$- and $D$-waves are accompanied by relativistic singlet and triplet $P$-waves.\,This can be seen explicitly in Fig.\,\ref{fig:bc} where the wave functions for the ground state of the $b\bar{c}$ mesons are depicted for the 4 types of mesons with quantum numbers $J^P=0^{\pm},1^{\pm}$, considered in this work. 

\section{Results and Discussion}
\label{results}  

\subsection{Mass spectra}
We consider 2 models in this work: model P1 was fitted to the masses of pseudoscalar states only, whereas model P2 was fitted to the masses of pseudoscalar, scalar, and vector mesons. The parameters of the models are listed in Table~\ref{tab:parameters}. Fitting the quark masses is much more time consuming than fitting the other parameters. Therefore, we first determined them in preliminary calculations and then held them fixed in the final fits of $\sigma$, $\alpha_s$ and $C$. This procedure is certainly good enough for the purpose of this work. Furthermore, 
early results clearly favored pure scalar+pseudoscalar confinement, so throughout this work we set $y=0$. Also the results turn out not to be very sensitive to the choice of the Pauli-Villars parameter $\Lambda$, so we set it to be $\Lambda=2 m_1$. Our results are given in Table~\ref{tab:masses} for the $b\bar{b}$, $b\bar{c}$, $b\bar{s}$, $b\bar{q}$, $c\bar{c}$, $c\bar{s}$, $c\bar{q}$ states where $\bar q=\bar u$ or $\bar d$.

It is well known that the Lorentz structure of a kernel determines the spin-dependent interactions, and it is certainly one of the attractive features of a covariant formalism that they are not treated perturbatively but on an equal footing with the spin-independent interactions. But in a general fit to all types of mesons one cannot really test the predictive power of the covariant kernels in this regard because all interactions are fitted simultaneously. 

What is remarkable is that a fit to a few pseudoscalar meson states only, which is insensitive to spin-orbit and tensor forces and which do not allow to separate the spin-spin from the central interaction, leads to essentially the same model parameters as a more general fit (the $rms$ between model P1 and P2 differ only by 6 MeV). This demonstrates that the covariance of the chosen interaction kernel is responsible for the very accurate  prediction of the spin-dependent quark-antiquark interactions \cite{Leitao:2016bqq}.

Besides models P1 and P2, and in order to investigate the role of the confining interaction, we tested a third model, PCoul, where we switch off the confining interaction and fit the data just with a Coulomb and a constant term. The $rms$ is significantly larger but some interesting observations can be made concerning the wave functions, to be presented in the next subsection.

\begin{table}[h!]
\caption{Kernel parameters of models P1 and P2. Both models use the quark masses $m_b=4.892$ GeV, $m_c=1.600$ GeV, $m_s=0.448$ GeV, and $m_u=m_d\equiv m_q=0.346$ GeV.}
\begin{center}
\begin{tabular}{|l|ccccc|}
\hline
Model & $\sigma$ [GeV$^2$] & $\alpha_s$ & $C$ [GeV] &  number of states used in fit & $rms$ [GeV]\\
\hline
P1      & 0.2493 & 0.3643 & 0.3491 & 9& 0.036\\
P2 	& 0.2247 & 0.3614 & 0.3377 & 25 &0.031 \\
\hline
PCoul 	& - & 0.5323 & 0.1264 & 9 &0.209 \\
\hline
\end{tabular}
\end{center}
\label{tab:parameters}
\end{table}

\begin{table}[h!]
\centering
\caption{Comparison of the mass spectra of all mesonic experimental states (with at with least one $b$ or $c$ quark content) and quantum numbers: $J^P$, $J=0,1$, $P=\pm$  and the theoretical mass predictions of model $P_1$ and $P_2$. The $\triangle$ and $\square$ symbols represent the states used in the fit P1 and P2, respectively. The states with no symbol assigned are pure predictions. All the masses are given in units of GeV. There is weak evidence (at 1.8 $\sigma$) that the $\Upsilon(1D)$ ($10.15$ GeV, marked with "?")  has been seen \cite{CLEO,BABAR}.}
\label{tab:masses}       
\begin{tabular}{llllllll}
&$J^P=0^-$ & \hspace{1.5cm}   &   $J^P=1^-$& \hspace{1.7cm}  & $J^P=0^+$ & \hspace{1.7cm}   &  $J^P=1^+$ 
\end{tabular}
\begin{tabular}{|l|lll|lll|lll|lll|}
\hline
&exp. & P1 & P2 & exp. & P1 & P2& exp. & P1 & P2 & exp. & P1 & P2\\  \hline
& 9.398$^{\triangle, \square}$ & 9.386 & 9.415 & 9.460$^{\square}$ & 9.470 & 9.487 & 9.859$^{\square}$ & 9.856 &9.850 & 9.896 & 9.886 & 9.875 \\
& 9.999$^{\triangle, \square}$ & 9.982 & 9.968 & 10.02$^{\square}$ & 10.02 & 10.00 & 10.23$^{\square}$ & 10.25 & 10.22 & 10.26 & 9.890 & 9.879 \\
$b\bar{b}$& 10.30 & 10.37 & 10.33 & 10.15(?) & 10.16 & 10.13 & - & 10.57 & 10.52 & 10.51 & 10.27 & 10.24 \\
& - & 10.68 & 10.63 & 10.36$^{\square}$ & 10.40 & 10.35 & - & 10.86 & 10.80 & - & 10.28 & 10.24 \\
& - & 10.96 & 10.89 & - & 10.49 & 10.44 & - & 11.13 & 11.03 & - & 10.60 & 10.54 \\
& - & 11.28 & 11.16 & 10.58$^{\square}$ & 10.71 & 10.65 & - & 11.48 & 11.32 & - & 10.60 & 10.54 \\ \hline
&6.275$^{\triangle, \square}$ & 6.302 & 6.319 & - & 6.394 & 6.397 & - & 6.745 & 6.730 & - &6.777 & 6.757  \\
$b\bar{c}$& 6.842 & 6.888 & 6.865 & - & 6.941 & 6.912 & - & 7.161 & 7.121 & - & 6.777 & 6.758 \\
& - & 7.293 & 7.246 & - & 7.057 & 7.019 & - & 7.505 & 7.445 & - & 7.191 & 7.146 \\\hline
& 5.367$^{\triangle, \square}$ & 5.362 & 5.367 & 5.415$^{\square}$ & 5.442 & 5.436 & - & 5.784 & 5.763 & 5.829 & 5.796 & 5.770 \\
$b\bar{s}$& - & 5.938 & 5.910 & - & 5.993 & 5.957 & - & 6.208 & 6.163 & - & 5.811 & 5.785 \\
& - & 6.349 & 6.297 & - & 6.093 & 6.051 & - & 6.559 & 6.495 & - & 6.234 & 6.184 \\ \hline
& 5.279$^{\triangle, \square}$ & 5.288 & 5.293 & 5.325$^{\square}$ & 5.366 & 5.360 & - & 5.709 & 5.688 &5.726 & 5.716 & 5.690 \\
$b\bar{q}$& - & 5.864 & 5.835 & - & 5.918 & 5.882 & - & 6.132 & 6.087 & - &5.735 & 5.708 \\
& - & 6.274 & 6.221 & - & 6.017 & 5.974 & - & 6.483 & 6.418 & - &6.157 & 6.106 \\ \hline
& 2.984$^{\triangle, \square}$ & 3.009 & 3.030 & 3.097$^{\square}$ & 3.110 & 3.120 & 3.415$^{\square}$& 3.424 & 3.424 & 3.518 & 3.461 & 3.454 \\
$c\bar{c}$& 3.639$^{\triangle, \square}$ & 3.647 & 3.627 & 3.686$^{\square}$ & 3.702 & 3.677 & 3.918 & 3.930 & 3.894 & - & 3.474 & 3.465 \\
& - & 4.123 & 4.073 & 3.773$^{\square}$ & 3.784 & 3.756 & - & 4.355 & 4.291 & - & 3.950 & 3.911 \\ \hline
& 1.968$^{\triangle, \square}$ & 1.944 & 1.966 & 2.112$^{\square}$ & 2.107 & 2.109 & 2.318$^{\square}$ & 2.399 & 2.396 & 2.459 & 2.434 & 2.422 \\
$c\bar{s}$& - & 2.612 & 2.591 & - & 2.697 & 2.667 & - & 2.910 & 2.872 & 2.535 & 2.458 & 2.444 \\
& - & 3.100 & 3.048 & - & 2.769 & 2.737 & - & 3.340 & 3.274 & - & 2.934 & 2.893 \\ \hline
& 1.867$^{\triangle, \square}$ & 1.858 & 1.881 & 2.009$^{\square}$ & 2.029 & 2.030 & 2.318$^{\square}$ & 2.319 &2.316 & 2.421 & 2.351 & 2.339 \\
$c\bar{q}$& - & 2.529 & 2.507 & - & 2.617 & 2.587 & - & 2.828 & 2.790 & - & 2.377 & 2.362 \\
& - & 3.016 & 2.964 & - & 2.687 & 2.655 & - & 3.257 & 3.191 & - & 2.852 & 2.810 \\ \hline
\end{tabular}
\end{table}

\subsection{Wave functions}
\begin{figure}[h!]
\centering
\includegraphics[width=10cm,clip]{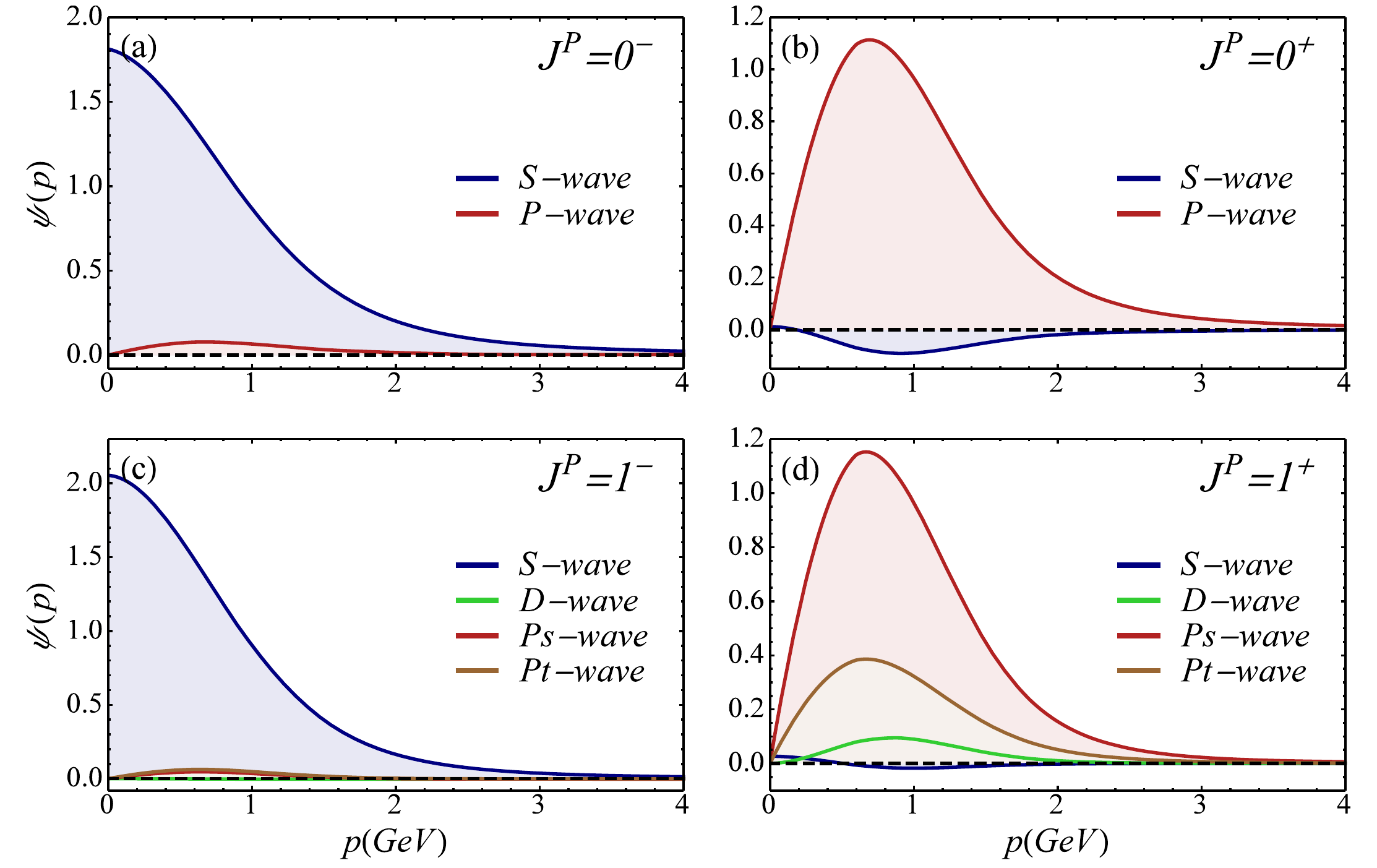}
\caption{Model P1 ground-state predictions for $b\bar{c}$ states.}
\label{fig:bc}       
\end{figure}

\begin{figure}[h!]
\centering
\includegraphics[width=13cm,clip]{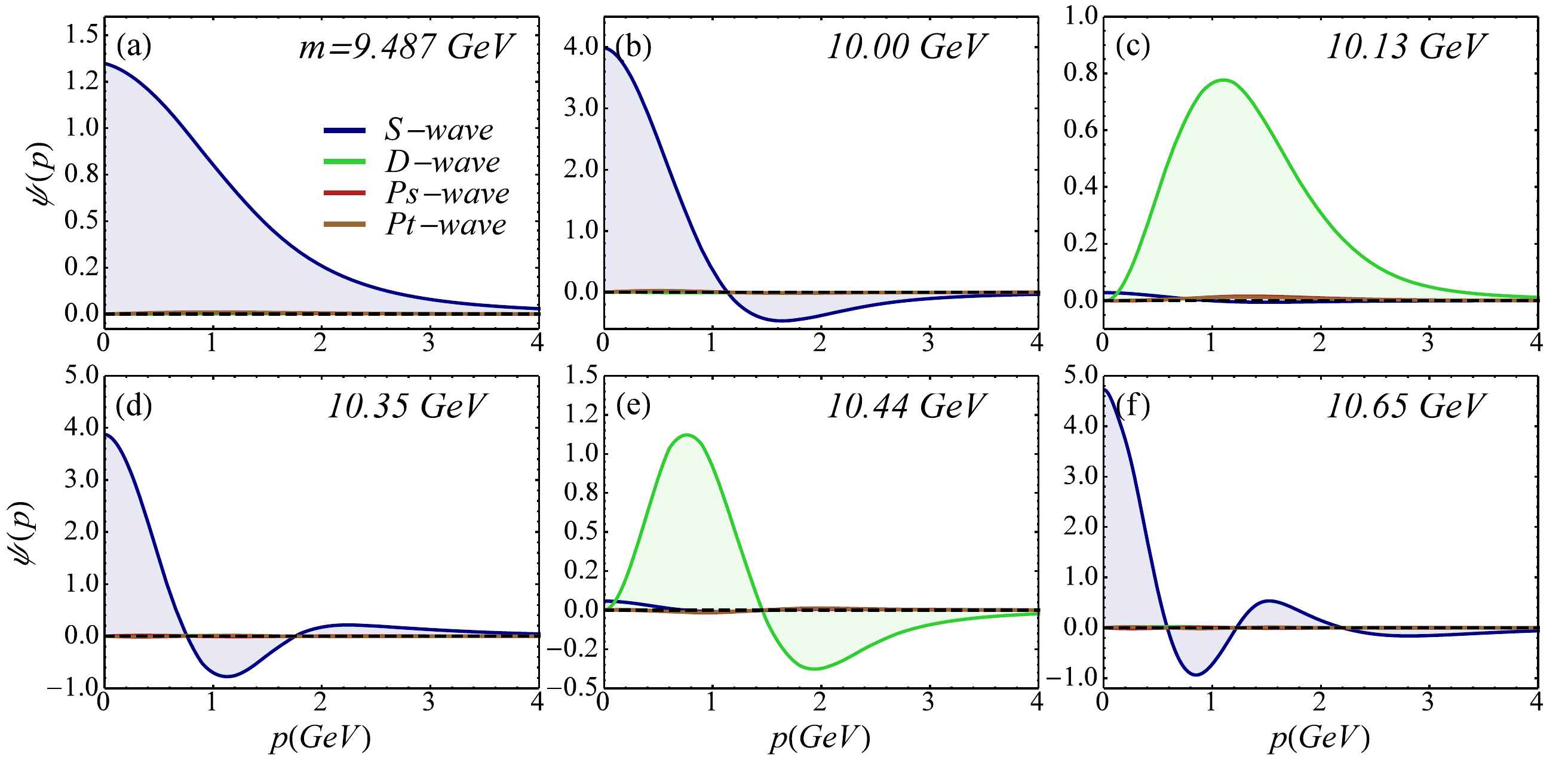}
\caption{Normalized wave-function components for the ground-state [Fig. (a)] and a few radial excitations for the bottomonium system with $J^P=1^-$ obtained with model P2.}
\label{fig-1}       
\end{figure}

\begin{figure}[h!]
\centering
\includegraphics[width=13cm,clip]{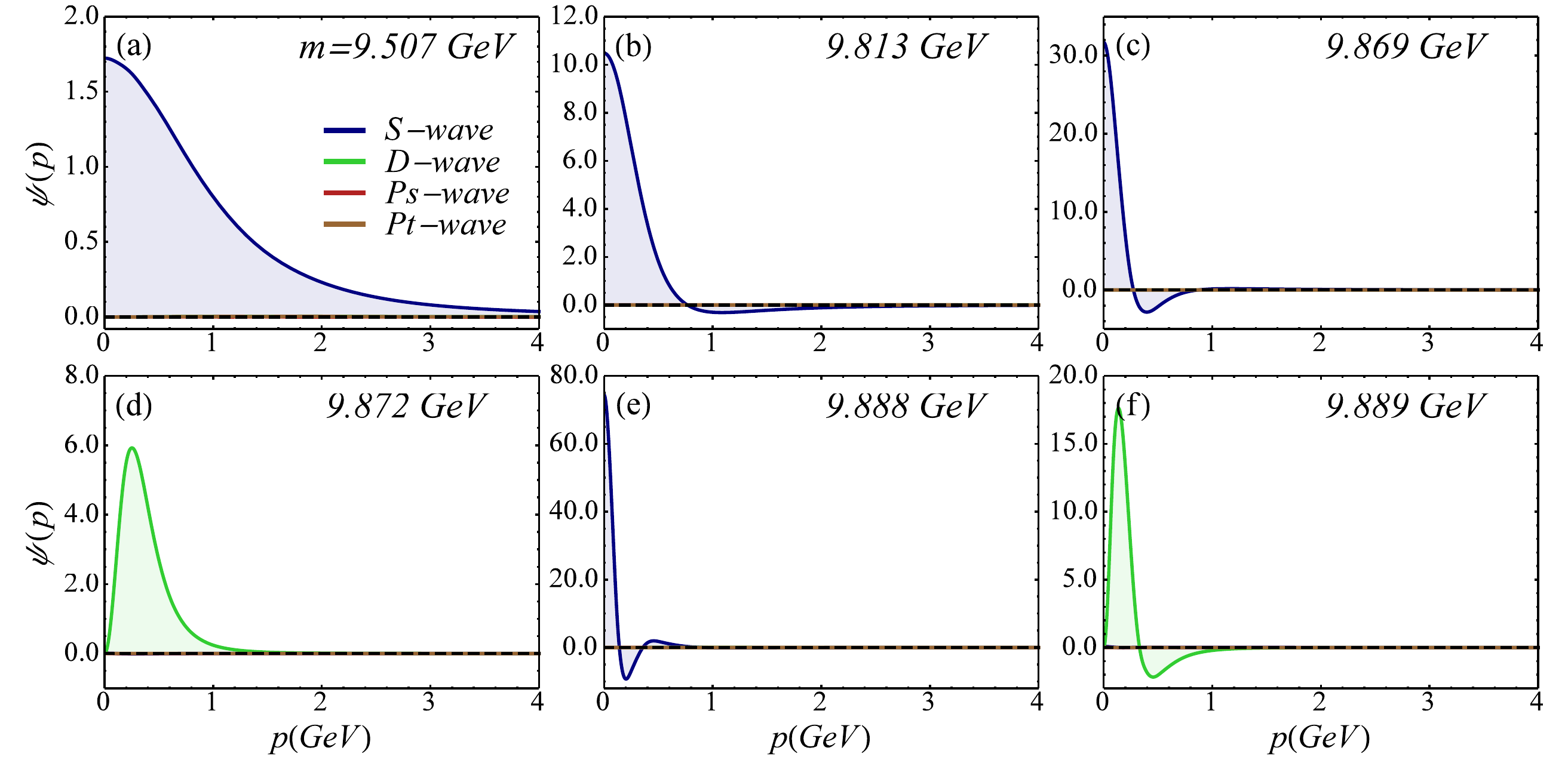}
\caption{Normalized wave-function components for the ground-state [fig.(a)] and a few radial excitations for the bottomonium system with $J^P=1^-$ obtained a toy model (PCoul) without confinement.}
\label{fig-2}       
\end{figure}

In Fig.\,\ref{fig:bc}, the wave function components of $S$-, $D$- and $P$- waves [$P_s$ (singlet)  and $P_t$ (triplet)]  of the ground-state of the $b\bar{c}$ system are given for the different types of mesons. These scalar functions are what we defined as $\psi_{j}^\rho(p)$  in Eq.~(\ref{eq:PSI}) and they are normalized as

\begin{align}
\int dp p^2 \left(\psi^2_S(p)+\psi^2_D(p) \right)=1, \qquad \text{for } J^P=0^{\pm},\\
\int dp p^2 \left(\psi^2_S(p)+\psi^2_D(p)+\psi^2_{Ps}(p)+\psi^2_{Ps}(p) \right)=1, \qquad \text{for } J^P=1^{\pm}.
\end{align}
By inspection we see that the relativistic components aforementioned are not completely negligible even for the $b\bar{c}$ state, usually assumed to be a nonrelativistic system.\,In Fig.\,\ref{fig-1} we depict the normalized wave function components for the bottomonium vector state using the predictions of model P2. What one observes is a pattern where the S-wave (depicted in navy-blue) always dominates, apart from the 2$^{nd}$ and the 4$^{th}$ radial excitations, where the D-wave component (marked in green) is the most prominent. 
However, for same predictions for the bottomonium wave functions made with model PCoul (see Fig.\,\ref{fig-2}) there is a change in the observed pattern, now the dominant  D-wave component appears to be in 3$^{rd}$ and the 5$^{th}$ radial excitations.
This is interesting because if a $\Upsilon(2D)$ state would be detected  it would be sensitive to the choice this potential. However, and just for what has been observed so far, a linear confining piece in the kernel seems to be necessary to get the ordering of the levels right.

%

\section{Summary and Outlook}
\label{summary}  

In this work we report on the recent developments of CST-BS formalism applied to heavy and heavy-light mesons.\,A very accurate  mass spectrum is obtained with just a few parameters. Furthermore, we observe that a fit  without any {\it direct} information about the spin-orbit and tensor forces (model P1) leads essentially to the same predictions for the model parameters as another fit (model P2) that takes into account i) more states and ii) states with explicit dependence on those forces. We have therefore shown that covariance leads to an accurate prediction of the spin-dependent quark-antiquark interactions. We have also checked that the radial excitations of the vector bottomonium can indirectly constrain the type of interaction potential chosen. 

In the near future we plan to calculate the mass spectrum of the tensor mesons. In the next step we will extend the formalism to the light meson sector in a consistent way by solving both the CST-Dyson (mass-gap equation), and the 4CSE. With all the covariant wave functions computed within this approach, it is then relatively straightforward to compute other observables such as radiative decays, other decay rates and calculations involving the structure of the $q\bar{q}$ states, as for instance electromagnetic and transition form factors.

\end{document}